\documentclass[aps,reprint,nofootinbib]{revtex4-1}

\usepackage[utf8]{inputenc}
\usepackage{amsfonts}
\usepackage{color}

\usepackage{amssymb}
\usepackage{amsmath}
\usepackage{stmaryrd}
\usepackage{latexsym}

\usepackage{xcolor}
\definecolor{myurlcolor}{HTML}{08457E}
\definecolor{mylinkcolor}{HTML}{2A52BE}
\definecolor{mycitecolor}{HTML}{E30022}

\usepackage{float}
\usepackage[colorlinks, linkcolor=mylinkcolor, citecolor=mycitecolor, urlcolor=myurlcolor, linktocpage=true]{hyperref}

\def\equationautorefname~#1\null{(#1)\null}
\def\tableautorefname~#1\null{(#1)\null}
\def\figureautorefname~#1\null{(#1)\null}
\def\sectionautorefname~#1\null{(#1)\null}

\let\origref\autoref
\def\autoref#1{\textbf{\origref{#1}}}

\let\origcite\cite
\def\cite#1{\textbf{\origcite{#1}}}

\usepackage{multirow}
\usepackage{tikz}
\usepackage{pgfplots}

\usepackage{titlesec}
\titleformat*{\section}{\centering\small\bfseries\scshape}
\titleformat*{\subsection}{\small\bfseries\scshape}
\titleformat*{\subsubsection}{\small\bfseries\scshape}

\newcommand{\be}{\begin{equation}}
\newcommand{\ee}{\end{equation}}
\newcommand{\bea}{\begin{eqnarray}}
\newcommand{\eea}{\end{eqnarray}}
\newcommand{\benn}{\begin{eqnarray*}}
\newcommand{\eenn}{\end{eqnarray*}}

\def\bse{\begin{subequations}}%
\def\ese{\end{subequations}}%

\def\L{{\cal L}}      
\def\S{{\cal S}}  
\def\R{{\cal R}}     

\newcommand{\sfrac}[2]{\dfrac{\,#1\,}{\,#2\,}}

\newcommand{\der}[2]{\sfrac{d #1}{d #2}}

\let\oldsqrt\sqrt
\def\sqrt{\mathpalette\DHLhksqrt}
\def\DHLhksqrt#1#2{%
	\setbox0=\hbox{$#1\oldsqrt{#2\,}$}\dimen0=\ht0
	\advance\dimen0-0.4\ht0
	\setbox2=\hbox{\vrule height\ht0 depth -\dimen0}%
	{\box0\lower0.4pt\box2}}

\begin{document}

\title{A Neutron Star with a Strange Quark Star-like Mass-Radius Relation}

\author{A. Sava{\c s} Arapo{\u g}lu}
\email{arapoglu@itu.edu.tr}

\author{A. Emrah Y{\"u}kselci}
\email{yukselcia@itu.edu.tr}
\affiliation{Department of Physics, Istanbul Technical University, 34469 Maslak, Istanbul, Turkey}

\begin{abstract}
We study the structure of spherically symmetric and static objects in the presence of a nonminimally coupled scalar field having a potential of the form $V(\phi)=-\mu^2\phi^2/2+\lambda\phi^4/4$. We numerically solve equations of the system using two different realistic equations of state for neutron matter and give the mass-radius relations as well as the radial profiles of the scalar field, densities and the mass function for a sample configuration. We show that a specific solution type encountered in such systems causes characteristic neutron star mass-radius relations to turn into the ones belonging to strange quark stars. 
\end{abstract}

\maketitle

\raggedbottom

\section{INTRODUCTION}

Scalar fields are frequently get involved in alternative gravitational theories with a potential and they may couple to either gravitational or matter sector depending on the choice of frame \cite{faraonibook}. Taking advantage of the ``arbitrariness" in the parameters determining the shape of the potential and the coupling bring about unexpected phenomena occurring around compact astrophysical objects such as spontaneous scalarization \cite{damour1992,damour1993,damour1996} that may drastically change the inner structure of an object, and screening mechanisms \cite{vainsthein1972,Khoury2004PRL,Hinterbichler2010} which hide their effects in the vicinity of a massive body while show the impacts at long distances. In addition to the restrictions coming from fundamental theories, this ``arbitrariness" can be constrained through the observations, for instance, by the help of the observational data obtained from solar system tests \cite{ppn_beta_1,ppn_beta_2,ppn_gamma,ppn_review}, dipole radiation in pulsar–white-dwarf binary systems \cite{dipole_rad_Freire:2012,dipole_rad_Shao:2017} and gravitational waves \cite{LIGO_GW170817,GW170817_observe}, etc.

Compact astrophysical objects constitute a class of natural laboratories to constrain the alternative gravitational theories \cite{Eksi2014}. In particular neutron stars are probably one of the most robust candidates to test models in strong gravitational regimes \cite{Stairs2003,psaltis2008}. Although their inner structure is not exactly known due to obscurity in their equation of state (EOS), one is still allowed to obtain related restrictions using the universality relations \cite{yagi2017rev} or even just the maximum mass observations \cite{Cromartie2019} by comparing the predictions of mass-radius (M-R) curves of the model at hand. In spite of the fact that there are many models of hypothetical compact objects in the literature \cite{oscillaton1,oscillaton2,oscillaton3,oscillaton4,Brito2015}, among them the ones called strange quark stars (SQS), which are also thought to may exist in the core of neutron stars \cite{weber2005rev,2012_Weber}, are closely related to the results of this study.

In this work the effects of nonminimally coupled scalar field on the structure of a spherically symmetric and static body are investigated by using a potential of the form $V(\phi)=-\mu^2\phi^2/2+\lambda\phi^4/4$, which was used before to define the symmetron field \cite{Hinterbichler2010} in the Einstein frame. The aim of this paper is to examine the results of a different solution type than the one encountered in screening mechanisms (e.g. caused by the symmetron field), outcomes from which differ due to initial conditions of the scalar field together with the definition of energy scales of the model. The differential equation system obtained from the field equations is solved numerically with two different EOS for neutron matter. It is shown that characteristic M-R relations of neutron stars turn into SQS M-R curves when the energy scales of the model is not specified accordingly, for instance contrary to Ref.\ \cite{Hinterbichler2010}, and the scalar field is unable to reach its vacuum expectation value due to inadequate size of the radius so that the object is unscreened or partially screened \cite{Sakstein2015}. (Although the Jordan frame formulation is taken into account here, the ``screening" term will be kept throughout this paper.) Another way to look at the result is that it is possible to mimic the M-R relations of SQS with the effect of macroscopic tools such as scalar fields rather than microscopic calculations of EOS coming from, for instance, MIT bag model \cite{MIT01,MIT02,MIT03}.

The investigated solution type describes an oscillating tail of the scalar field outside the star and it is commonly encountered in the configurations with scalar cores \cite{Brito2015}. Although in these solutions the metric functions describe the Minkowskian spacetime asymptotically, periodic oscillations occur in the defined mass function. Similar situation is known for time-dependent objects called oscillatons for which the amplitude of the oscillations should be negligible in order to obtain a proper model \cite{oscillaton1,oscillaton2,oscillaton3,oscillaton4}.

The plan of the paper is as follows : In Sec.\ \autoref{setup} we derive the field equations and define the differential equation set that will be solved numerically. In Sec.\ \autoref{model} we give the model in detail including the comparison with the literature, and the results in Sec.\ \autoref{analysis} followed by the final comments in Sec.\ \autoref{conclusion}.

\section{HYDROSTATIC EQUILIBRIUM} \label{setup}
We consider the following action written in the Jordan frame
\begin{equation}
\begin{aligned}
	\S = \int \! d^4 x \, \sqrt{|g|} \, \bigg[ &\sfrac{1}{2 \kappa} \big( 1 + \kappa \xi \phi^2 \big) \R \\ &- \sfrac{1}{2} \nabla^c \phi \, \nabla_{\!\!c} \, \phi - V\!(\phi) + \L^{(m)} \bigg] \;.
\end{aligned}
\label{eq:action}
\end{equation}
where $\xi$ is the coupling constant and $\L^{(m)}$ is the matter Lagrangian. We use geometrical units ($G = c = 1$) so that $\kappa = 8\pi G = 8\pi$. Equations of motion from the above action are obtained in the following form
\begin{subequations}
\begin{align}
	 \R_{\mu \nu} - \sfrac{1}{2} \R \, g_{\mu \nu} &= \, \kappa_{\rm eff} \big[ T_{\mu\nu}^{(m)} + T_{\mu\nu}^{(\phi)} \big] \;, \label{eq:fieldeq1} \\[2mm]
	 \boxempty \phi &= - \der{V_{\rm eff}}{\phi} \;, \label{eq:fieldeq2}
\end{align}
\end{subequations}
where $V_{\rm eff}$, $\kappa_{\rm eff}$, $T_{\mu\nu}^{(m)}$ and $T_{\mu\nu}^{(\phi)}$ are defined as
\begin{subequations}
\begin{align}
    V_{\rm eff}(\phi) &= -V(\phi) + \sfrac{1}{2}\xi\R\phi^2 \label{eq:eff_pot} \;, \\[2mm]
	\kappa_{\rm eff}(\phi) &= \kappa \big( 1 + \kappa \xi \phi^2 \big)^{-1} \label{eq:kappa_eff} \;, \\[2mm]
	T_{\mu\nu}^{(m)} &= \big(\rho + P\big) u_\mu u_\nu + P g_{\mu\nu} \label{eq:matter_tensor} \;, \\[2mm]
	T_{\mu\nu}^{(\phi)} &= \nabla_{\!\!\mu} \, \phi \, \nabla_{\!\nu} \, \phi - g_{\mu \nu} \bigg[ \sfrac{1}{2} \, \nabla^c \phi \, \nabla_{\!\!c} \, \phi + \, V\!(\phi) \bigg] \nonumber \\ & \quad- \xi \big( g_{\mu \nu} \boxempty - \nabla_{\!\!\mu} \, \nabla_{\!\nu} \big) \phi^2 \: .\label{eq:scalar_tensor}
\end{align}%
\end{subequations}
This interpretation of the equations of motion ensures that the energy-momentum tensor of the fluid is conserved $\nabla^{\mu} T_{\mu\nu}^{(m)} = 0$ whereas $\nabla^{\mu} T_{\mu\nu}^{(\phi)} \neq 0$ \cite{faraonibook}.

We look for spherically symmetric and static solutions described by the metric
\begin{equation}
	ds^2 = - e^{2f(r)} dt^2 + e^{2g(r)} dr^2 + r^2 d\Omega^2 \;,
\label{eq:metric}
\end{equation}
where $f(r)$ and $g(r)$ are the metric functions of radial coordinate only. Applying this metric to Eqs.\ \autoref{eq:fieldeq1} and \autoref{eq:fieldeq2} we obtain the following set
\begin{subequations}
	\begin{align}
	g' &= \bigg[ \sfrac{1}{r} \!+\! \xi \phi \phi' \bigg]^{-1} \bigg[ \sfrac{1 \!-\! e^{2g}}{2r^2} \!+\! \kappa_{\rm eff} \bigg\{ \sfrac{1}{2} \big(\rho \!+\! V\big)e^{2g} \nonumber \\ & \qquad + \xi \phi \bigg( \phi'' \!+\! \sfrac{2\phi'}{r} \bigg) \!+\! \bigg(\xi \!+\! \sfrac{1}{4}\bigg)\phi'^2 \bigg\} \bigg] \label{eq:components_tt} \: , \\[3mm] 
	f' &= \bigg[ \sfrac{1}{r} \!+\! \xi \phi \phi' \bigg]^{-1} \bigg[ \sfrac{e^{2g} \!-\! 1}{2r^2} + \kappa_{\rm eff} \bigg\{ \sfrac{1}{2} \big(P \!-\! V\big)e^{2g} \nonumber \\ & \qquad + \sfrac{1}{4}\phi'^2 - \sfrac{2\xi\phi\phi'}{r} \bigg\} \bigg] \label{eq:components_rr} \: , \\[3mm] 
	f'' &= -\big(f'\!-\!g'\big) \Big[ f' \!+\! \sfrac{1}{r} \Big] \!+\! \kappa_{\rm eff} \bigg\{ \! \big(P\!-\!V\big)e^{2g} \nonumber \\ & \quad \!-\! 2\xi\phi \bigg[ \phi'' \!+\! \phi' \bigg( \! f'\!-\!g'\!+\!\sfrac{1}{r} \! \bigg) \! \bigg] \!-\! \bigg( \! 2\xi \!+\! \sfrac{1}{2}\bigg) \phi'^2 \bigg\} \label{eq:components_QQ}  \: , \\[3mm] 
	\phi'' &= -\bigg[f' \!-\! g' \!+\! \sfrac{2}{r}\bigg]\phi' + 2\xi\phi \bigg[f'' \nonumber \\ & \qquad \!+\! \big(f'\!-\!g'\big) \Big( f' \!+\! \sfrac{2}{r} \Big)  + \sfrac{1 \!-\! e^{2g}}{r^2}\bigg] + e^{2g} \der{V}{\phi} \label{eq:scalar_eq} \: , \\[3mm] 
	P' &= - f'\big(P + \rho\big) \label{eq:pressure_eq} \: ,
	\end{align}
\label{eq:thesystem}%
\end{subequations}
where first three equations are derived from $tt$, $rr$, and $\theta\theta$ components of Eq.\ \autoref{eq:fieldeq1}, respectively, and the last one is the result of the conservation equation, i.e. $\nabla^{\mu} T_{\mu\nu}^{(m)} = 0$. On the other hand, we use the following integral to compute the total mass of the configuration
\begin{equation}
	M \equiv \lim_{r \rightarrow \infty} m(r) = 4\pi \int_{0}^{\infty} r^2 E(r) \, dr
\label{eq:mass}
\end{equation}
where the function $E(r)$ is 
\vspace{2mm}
\begin{equation}
\begin{aligned}
	E(r) &= \sfrac{\kappa_{\rm eff}}{\kappa} \bigg[ \rho + \sfrac{1}{2} \big( \phi' e^{-g} \big)^2 + V(\phi) \\ & \quad + 2\xi \bigg\{ \phi \bigg( \! \phi'' + \phi' \bigg[ \sfrac{2}{r} \!-\! g' \bigg] \bigg) + (\phi')^2 \bigg\} e^{-2g} \bigg]
\label{eq:mass_E}
\end{aligned}
\end{equation}
as defined in Refs.\ \cite{salgado1998,Arapoglu2019}. The presence of the scalar field requires that the mass integral has to be evaluated at spatial infinity in order to consider its contribution to the total mass outside the configuration that is determined by the condition $P(R)=0$. This point arises in our case and is clarified in the next sections in detail.

\section{THE MODEL} \label{model}
As mentioned before we consider a potential of the form
\begin{equation}
	V(\phi) = - \sfrac{1}{2} \mu^2 \phi^2 + \sfrac{1}{4} \lambda \phi^4
\label{eq:potential}
\end{equation}
where $\mu^2>0$ and $\lambda>0$ are constants. For this configuration radial profile of the scalar field is governed by the effective potential defined in Eq.\ \autoref{eq:eff_pot} that yields
\begin{equation}
    V_{\rm eff}(\phi,\R) = \sfrac{1}{2} \omega \phi^2 - \sfrac{1}{4} \lambda \phi^4 \:,
\label{eq:sca_eff_pot}
\end{equation}
where $\omega\equiv\mu^2+\xi\R$. Critical points of this effective potential should be examined depending on the sign of $\omega$. For $\omega<0$ and $\omega=0$, there is only one critical point that is $\phi_{\rm cr}=0$ and is unstable. On the other hand, for $\omega>0$, the critical point $\phi_{\rm cr}=0$ becomes stable whereas the others, $\phi_{\rm cr}=\pm\sqrt{|\omega|/\lambda}$, are unstable. This means that shape of the effective potential changes depending on the radial profile of the star and the parameters of the model at hand.

Stellar configurations with a Higgs-like potential, which is equivalent to the potential considered here up to a constant, were first investigated in Ref.\ \cite{Fuzfa2013} where it was shown that a unique initial condition for the scalar field must be chosen so that its value reaches to the one of the unstable critical points of the effective potential in the exterior region and hereby asymptotic flatness is achieved. Recently, for the same model, the structure of neutron stars with various realistic EOS was studied in Ref.\ \cite{Arapoglu2019} and constraints on the parameters of the model based on observations were obtained with the analysis of the resulting M-R curves. For that scenario, on the contrary to unstable critical points, the stable critical point of the effective potential, i.e. $\phi_{\rm cr}=0$, gives asymptotically de Sitter space since $V_{\rm Higgs}(\phi=0)\neq0$. That is why one has to find the unique initial value for the scalar field which matches with the value of the unstable critical point at infinity. However, the asymptotic flatness can be achieved for the potential given in Eq.\ \autoref{eq:potential} because there is no ``problematic" constant term in the expression and, therefore, $V(\phi=0)=0$. Here we examine the structure of neutron stars in the context of the scalar field motion around the stable critical point of the effective potential, namely $\phi_{\rm cr}=0$.

In systems with those kind of potentials, a stable solution providing asymptotic flatness for an isolated stellar configuration requires a finely-tuned initial condition for the scalar field as mentioned above. In fact there is one and only numerical value for each object. In those cases the scalar field approximately decays with a Yukawa-type profile, $\sim exp(-\alpha r)/r$, outside the star \cite{Hinterbichler2010,Fuzfa2013,Fuzfa2014} and asymptotically reaches to a constant value that is the value of the unstable critical point of the effective potential and, therefore, the metric functions describe Minkowskian spacetime as $r\rightarrow\infty$. However, there are two more different kinds of solutions. In one group the metric functions and the radial profile of the scalar field have divergent solutions (For details see Fig.\ (4) in Ref.\ \cite{Arapoglu2019}) and they will not be discussed further here. The other group, which is the main concern of this paper, is valid for the initial conditions obeying $|\phi_c|<\sqrt{\omega/\lambda}$ and their resulting behavior is found to be damped oscillation around $\phi=0$ following $\sim \sin(\sqrt{\omega}r)/r$ approximately.

The potential \autoref{eq:potential} has been used to describe a screening mechanism realized by a scalar field called the symmetron which was originally proposed in the Einstein frame \cite{Hinterbichler2010}. Although the potential differs in two frames \cite{faraonibook}, namely the Jordan and the Einstein frames, it is useful to compare the outcomes due to the fact that the effect of the underlying mechanism on the objects stays the same, in general, in the context of the deviations from general relativity. The symmetron field defined in the Einstein frame designates a screening mechanism up to a certain energy scale that is the critical density of our universe today $\rho_{\rm crit}\sim H_0^2 M^2_{\rm pl}$ \cite{Hinterbichler2010,copeland2019}. Therefore, around a compact astrophysical objects the corresponding solution is always the one that decays with a Yukawa-type profile as referred before. However, if we consider energy scales for the scalar field in a general manner, then ``an object can be screened, unscreened or even partially screened" \cite{Sakstein2015}. The solution that shows damped oscillatory behavior around $\phi=0$ corresponds to unscreened cases and it will be investigated in the next section with its effects on the structure of the object.

\section{NUMERICAL ANALYSIS AND MASS-RADIUS CURVES} \label{analysis}

\begin{figure*}[!t]
	\centering

	\begin{tabular}{@{}c@{}}
		\includegraphics[width=.40\linewidth]{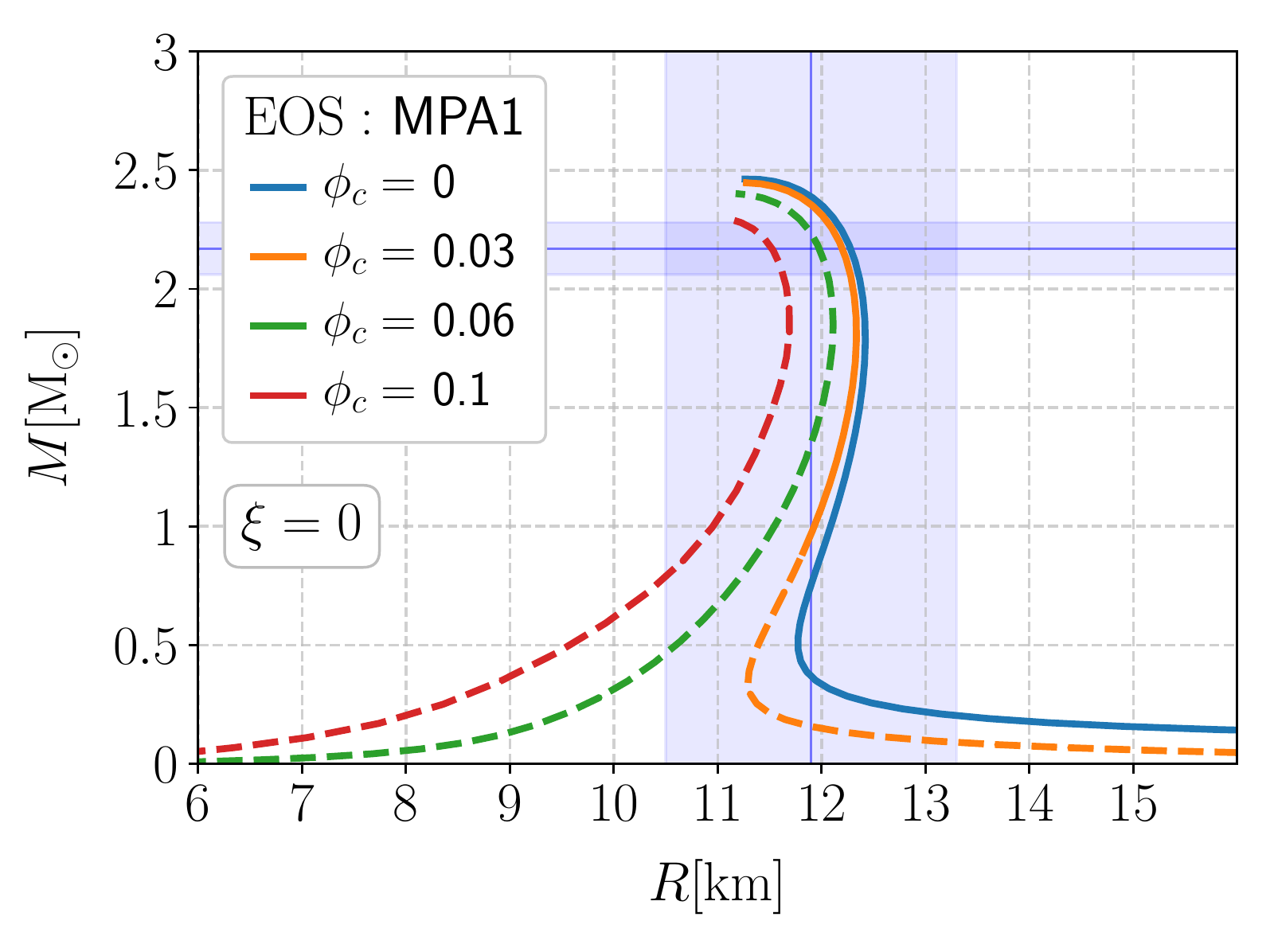}
		\label{fig:mr_mpa1_0}
	\end{tabular}\hspace{5mm}
	\begin{tabular}{@{}c@{}}
		\includegraphics[width=.40\linewidth]{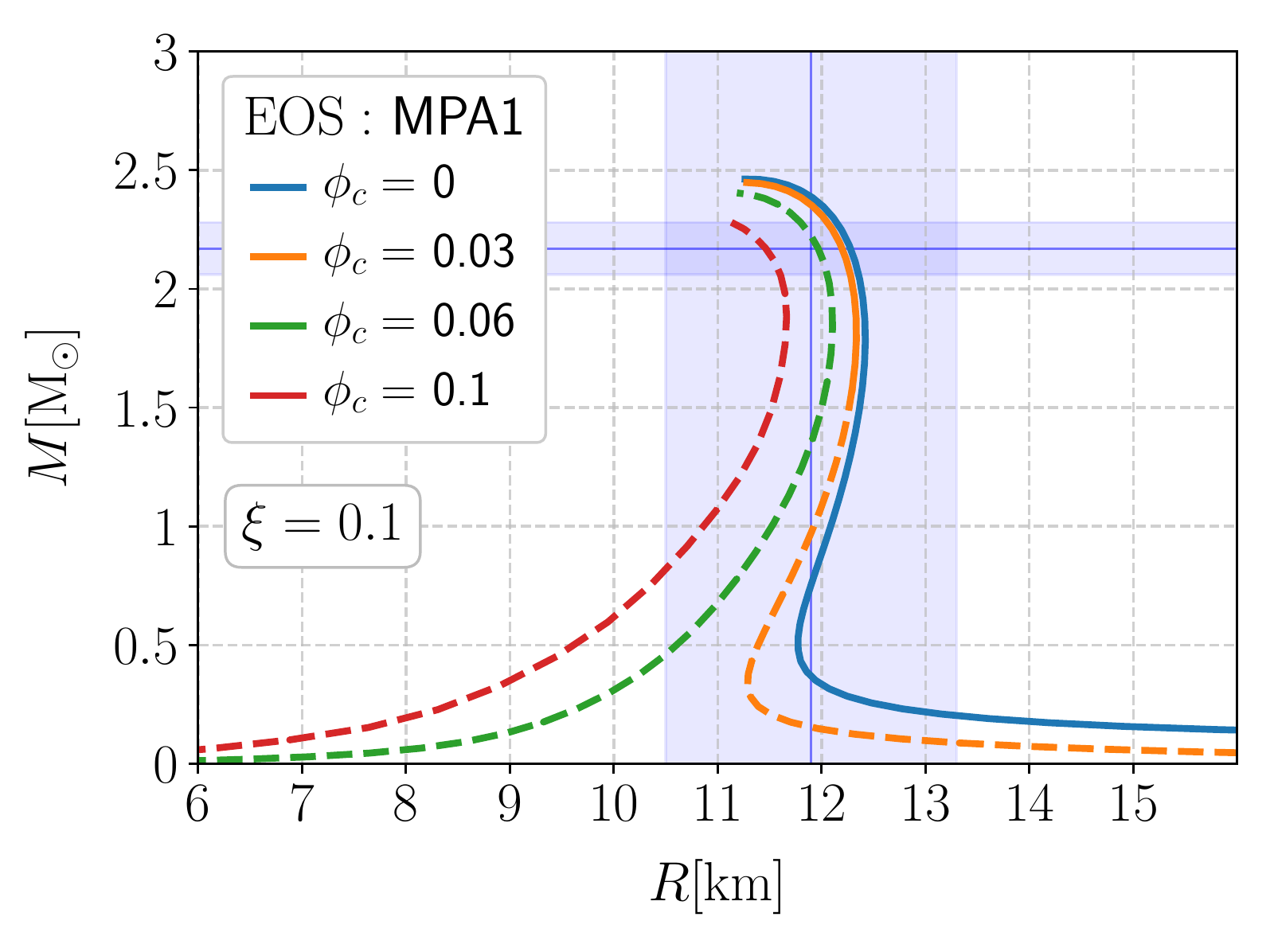}
		\label{fig:mr_mpa1_03}
	\end{tabular}
	
	\begin{tabular}{@{}c@{}}
		\includegraphics[width=.40\linewidth]{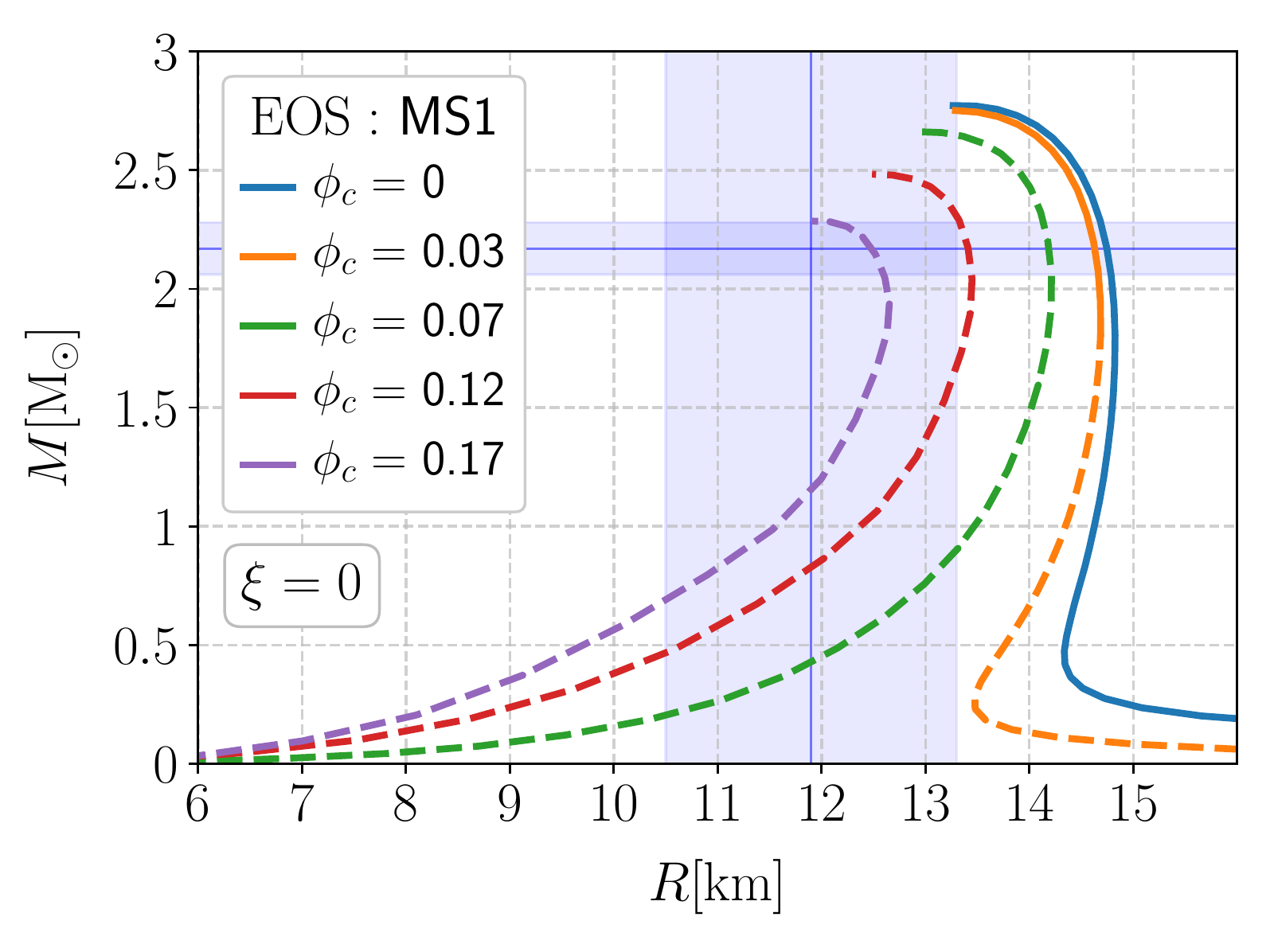}
		\label{fig:mr_ms1_0}
	\end{tabular}\hspace{5mm}
	\begin{tabular}{@{}c@{}}
		\includegraphics[width=.40\linewidth]{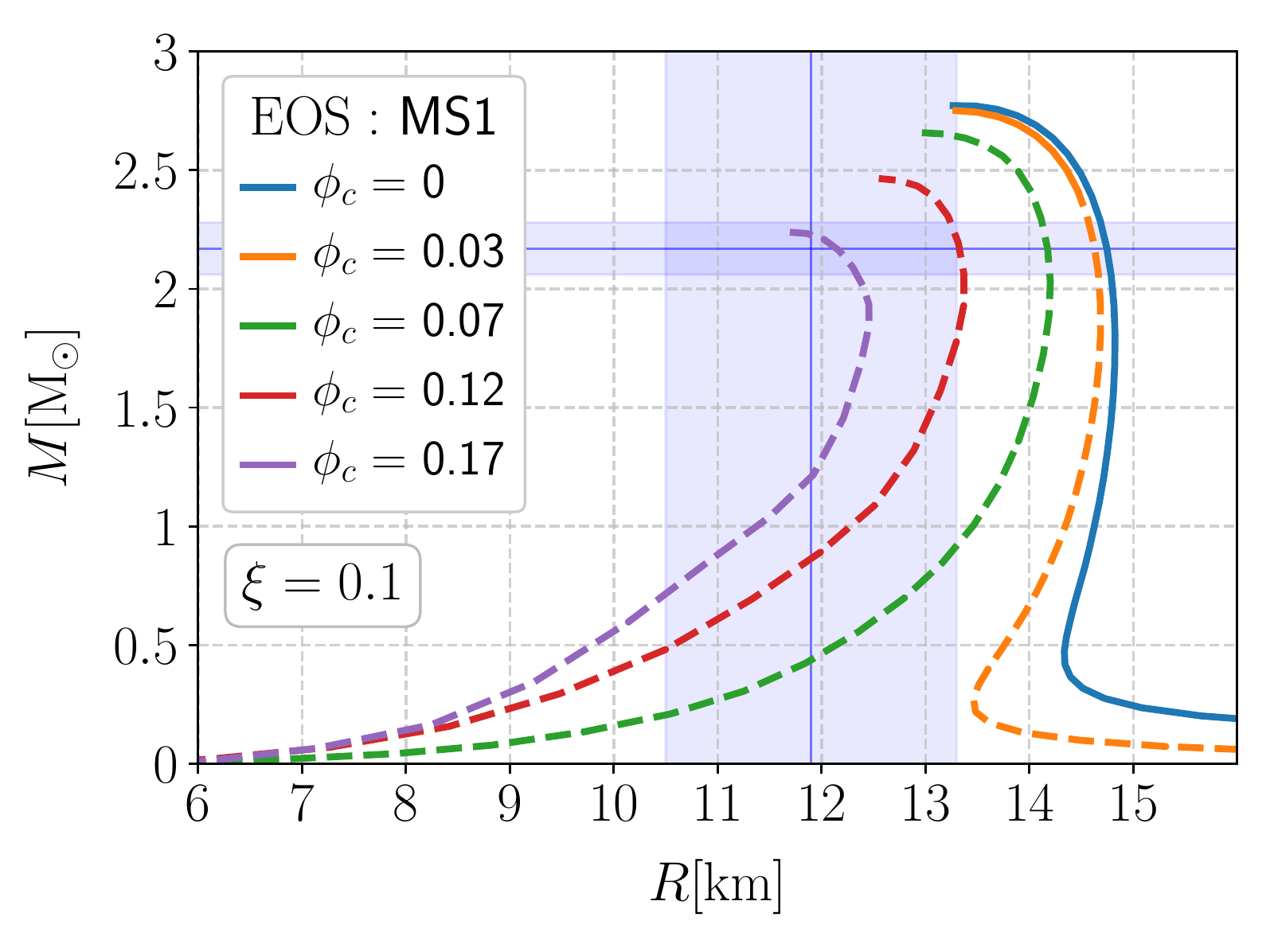}
		\label{fig:mr_ms1_01}
	\end{tabular}
	
	\caption{M-R relations for two different EOS (MS1 and MPA1) with various central values of the scalar field. Values of the coupling constant are $\xi=0$ and $\xi=0.1$ on the left and right panels, respectively. Dashed parts of the curves represent configurations with negative energy density at the inner layers due to the scalar field, for which a particular example is shown in Fig.\ \autoref{fig:quarky_funcs}. Here potential parameters are $\mu \!=\! \lambda \!=\! 1$ for simplicity. Horizontal solid line is $2.17^{+0.11}_{-0.10}\,\rm M_{\odot}$ for J0740+6620 \cite{Cromartie2019} whereas vertical solid line, $R \!=\! 11.9 \pm 1.4\,\rm km$, stands for the radius constraint obtained from GW170817 event \cite{LIGO_GW170817}.}
	\label{fig:mr_curves}
\end{figure*}

\subsection{Numerical Analysis}

The system given in Eq.\ \autoref{eq:thesystem} is solved numerically starting from the center of the configuration to spatial infinity, which, in view of computational convenience, is described as the distance from the center where the scalar field almost reaches its asymptotic value. Mass of the star, on the other hand, is computed by integrating the expression given in Eq.\ \autoref{eq:mass}. Two different realistic EOS, namely MS1 \cite{eos_ms1} and MPA1 \cite{eos_mpa1}, are used and included in the system through numerical interpolation. To initiate the integration the following boundary conditions are imposed on the system : $\phi(0)\!=\!\phi_{\rm c}$, $\phi'(0)\!=\!0$, $P(0)\!=\!P_{\rm c}$, $g(0)\!=\!0$, $M(0)\!=\!0$ considering regularity at the center and $g(\infty)\!=\!0$, $f(\infty)\!=\!0$ in order to satisfy the asymptotic flatness at spatial infinity. Therefore, there is no need to take an extra action for matching the interior and the exterior solutions at the radius of the star, $R$, which is determined by the condition $P(R)=0$ and the corresponding value of the mass function, $m(R)=M$, is taken as the total mass of the star.

At this point some further comments are required for determining the mass and the radius of a configuration in our model. Periodic oscillations in the mass function are encountered outside the star $(r>R)$ because of the investigated solution type which does not have an exponential suppression at spatial infinity. Although this raises a question about determination of the proper radius of the star since the contribution of the scalar field to the mass function seems to be continuing in the exterior region as well, these oscillations occur as periodic deviations around the mass value determined by the condition $P(R)=0$. Then, for our purposes it is acceptable to consider this condition in order to construct M-R curves. The crucial point here is that amplitude of the oscillations has to be small in comparison with the whole configuration so that their effects are negligible.

\begin{figure*}[htpb]
	\centering

	\begin{tabular}{@{}c@{}}
		\includegraphics[width=.38\linewidth]{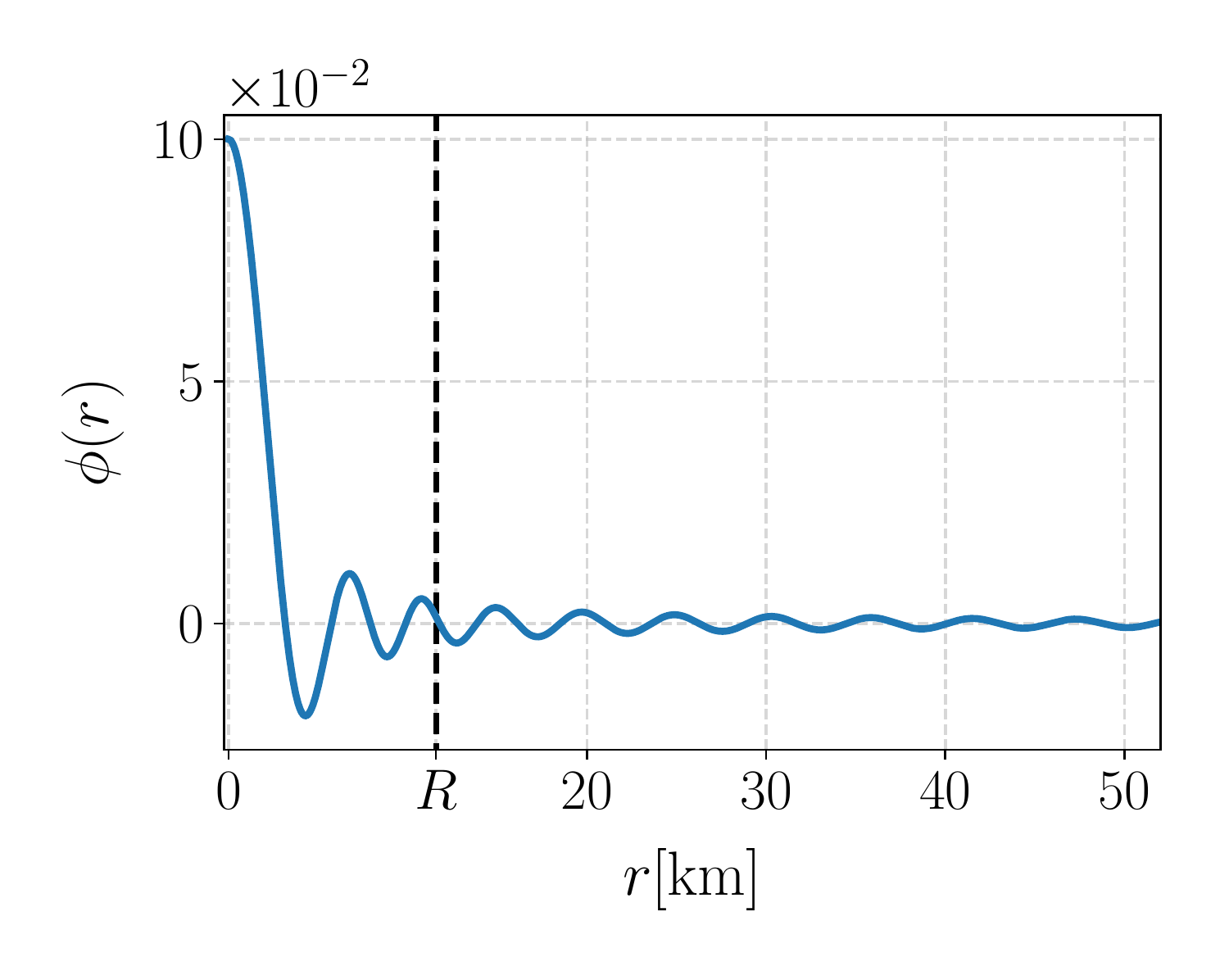}
	\end{tabular} \hspace{2mm}
	\begin{tabular}{@{}c@{}}
		\includegraphics[width=.38\linewidth]{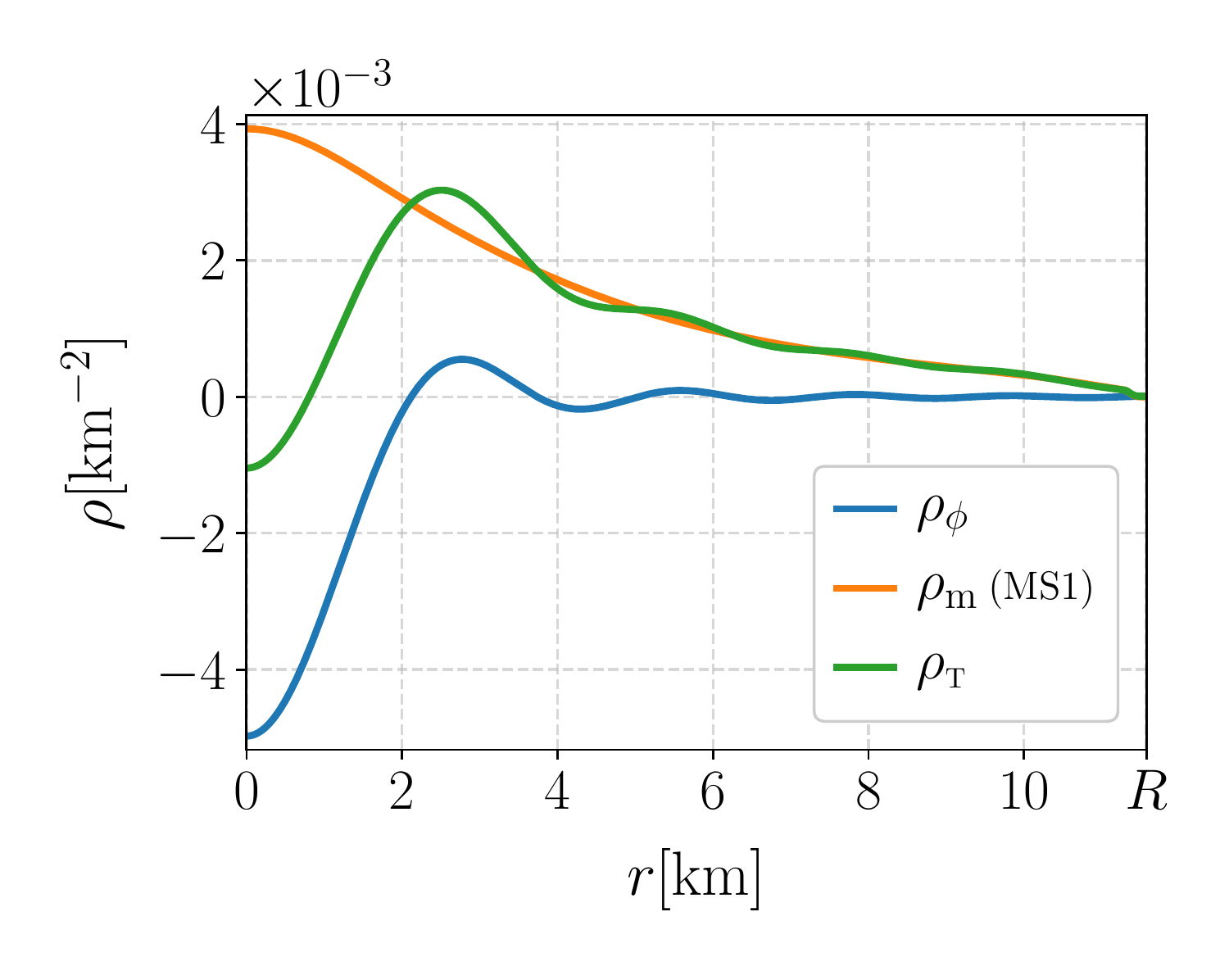}
	\end{tabular}
	
	\vspace{-4mm}
	
	\begin{tabular}{@{}c@{}}
		\includegraphics[width=.42\linewidth,trim={5mm 0 6.7mm 0},clip]{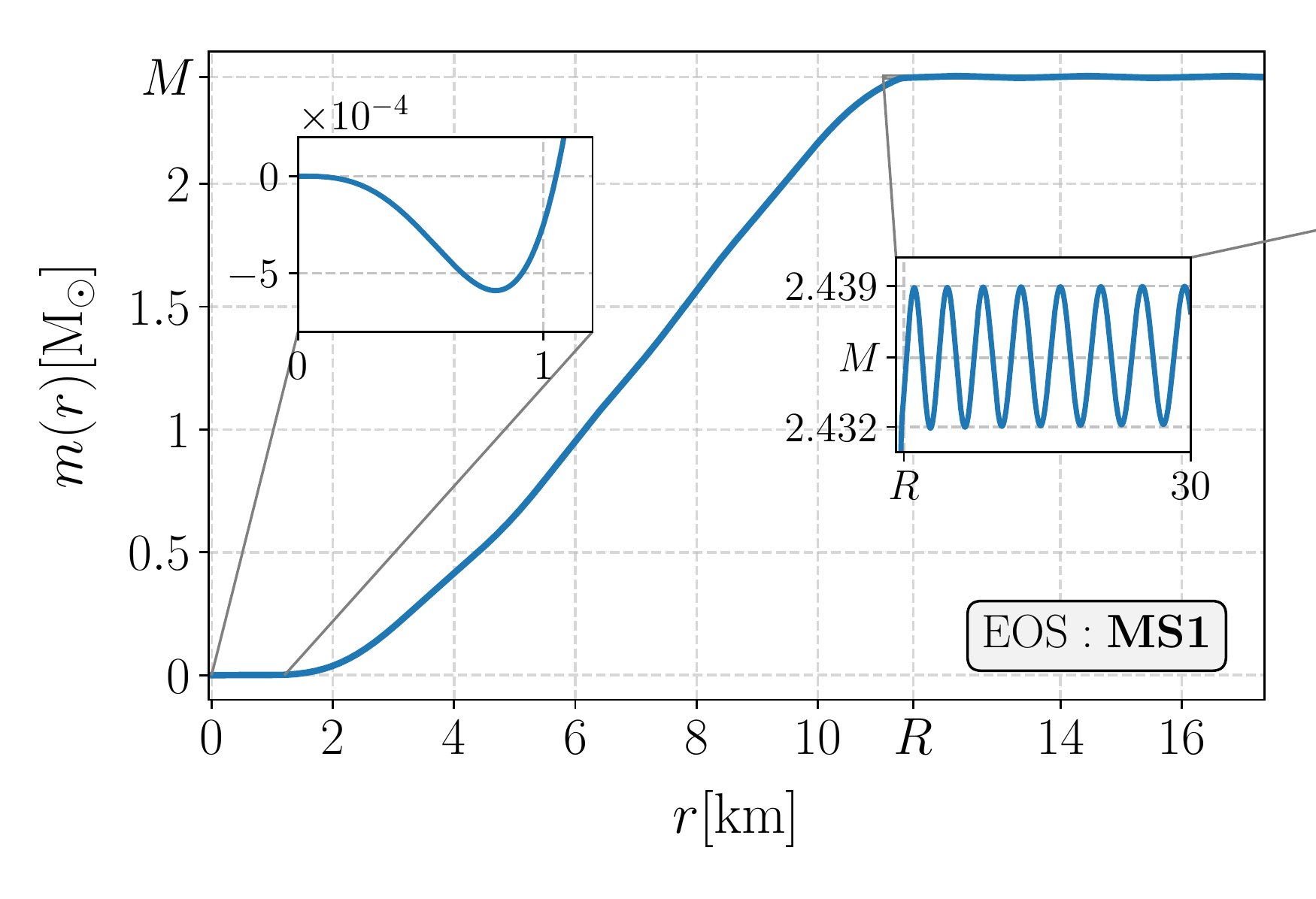}
	\end{tabular} \hspace{2mm}
	\begin{tabular}{@{}c@{}}
		\includegraphics[width=.395\linewidth]{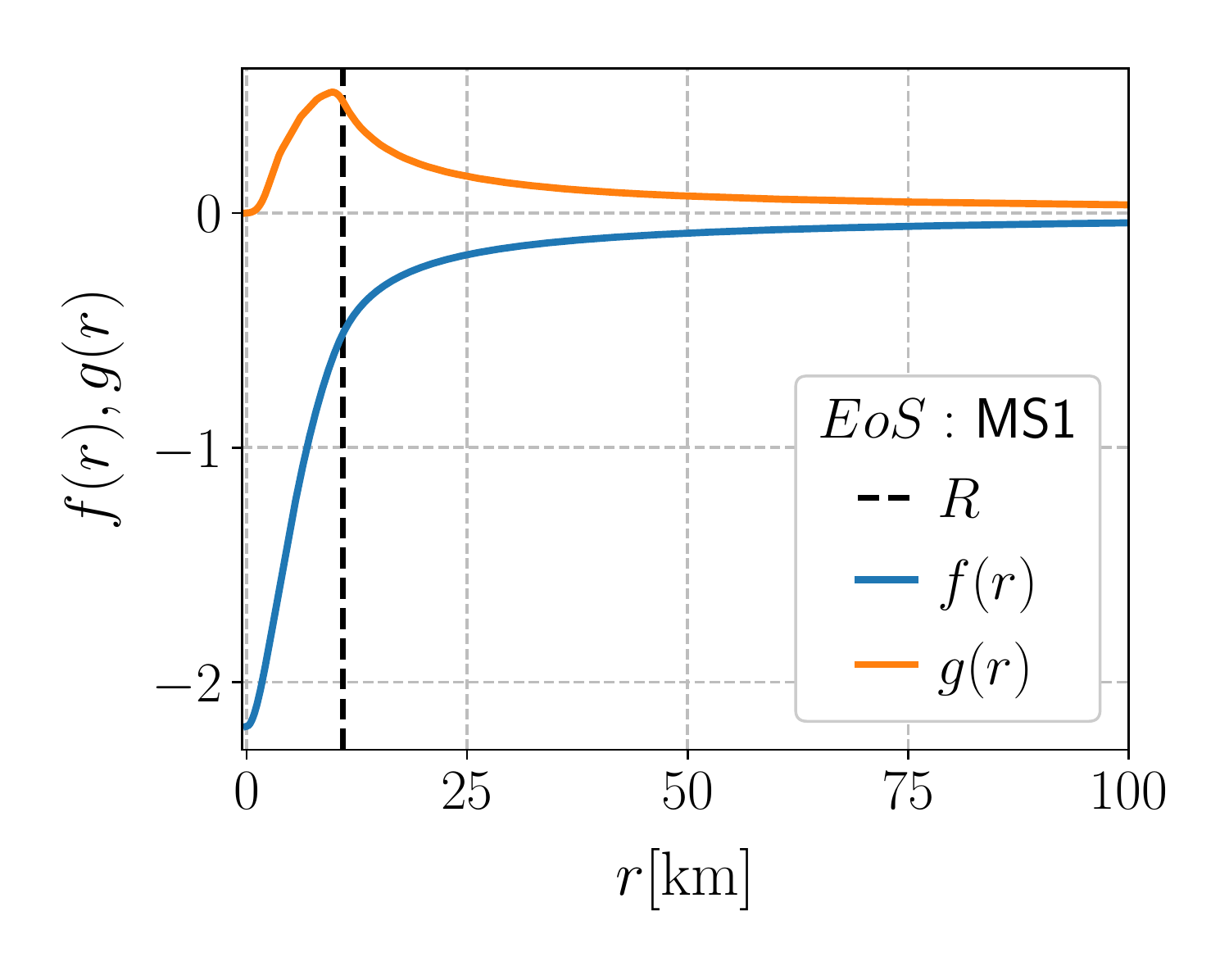}
	\end{tabular}
	
	\vspace{-4mm}
	
	\caption{Radial profiles of the scalar field (upper-left), densities (upper-right), the metric functions (lower-right) and the mass function (lower-left) for a sample configuration. Here $\phi_c=0.1$, $\xi=0$ and $\mu = \lambda = 1$ are chosen for simplicity. $R\approx11.575\,{\rm km}$ and $M\approx2.435\,{\rm M_{\odot}}$ indicate the radius and the mass of the star, respectively.}
	\label{fig:quarky_funcs}
\end{figure*}

In the light of above considerations, M-R relations are presented in Fig.\ \autoref{fig:mr_curves} with a sample of central value for the scalar field and two different values of the coupling constant $\xi=0$ (left) and $\xi=0.1$ (right). It is clear that these relations radically differ from the generic neutron star M-R curves and this result is due to effect of the potential and the chosen solution type on the stellar configuration since such an outcome is not obtained for zero or Higgs-like potentials for the same model \cite{Arapoglu2019}. This type of M-R relations encountered in SQS \cite{weber2005rev,2012_Weber,yagi2017rev} are obtained through the usage of EOS for strange quark matter based on MIT bag model \cite{MIT01,MIT02,MIT03}. Here the same characteristic M-R curves are acquired by using EOS for neutron matter and a scalar field with potential \autoref{eq:potential}.

The radial profiles of a sample configuration is given in Fig.\ \autoref{fig:quarky_funcs}. As can be seen in the figure domination of the scalar field over neutron matter in the core region of the star gives rise negative total energy density starting from the center up to a particular radius from which the neutron matter density becomes the dominant part. Dashed parts of M-R curves in Fig.\ \autoref{fig:mr_curves} indicate the presence of negative energy density at the core whereas this does not occur in the configurations represented by the solid parts. This negativity is proportional to the coupling constant, that is, if the value of $\xi$ is increased then the size of this region also increases for a specific configuration. Consequently, the number of this type of configurations in one particular M-R sequence grows with the increment of the coupling constant value as can be seen in Fig.\ \autoref{fig:mr_curves} by comparing the cases with $\xi=0$ and $\xi=0.1$. Additionally, amplitude of the oscillations in the mass function and the negativity at the core region grow in lower mass configurations. It is found that, e.g. for a configuration having a total mass of $0.5\,{\rm M_{\odot}}$, the former and the latter do not exceed $10\,\%$ of the total mass and $30\,\%$ of the radius, respectively. Furthermore, behavior of the scalar field is also responsible for the oscillations in the geometry, i.e. oscillations of the metric function $g(r)$, as pointed out in Ref.\ \cite{Brito2015} as well and similar to the results of Ref.\ \cite{Mahmoodzadeh2017}. Since we have constructed all solutions obeying the boundary conditions in accordance with the requirement of asymptotic flatness, these oscillations quickly fade away outside the star, an example of which can be seen in Fig.\ \autoref{fig:quarky_funcs} for the considered sample configuration.

Another important result is the fact that if the allowed maximum mass configuration of a particular M-R curve constitutes a core with negative energy density then this M-R curve shows exactly the same characteristics with bare SQS, while, if this is not the case, i.e. the maximum mass configuration has no negative energy density at the core, then the corresponding sequence tends to mimic M-R relations of SQS with nuclear crust (See Fig.\ (15) in Ref.\ \cite{weber2005rev}). Final point to report is that the parameter $\lambda$ has no significant effect on the shape of M-R curves, therefore, the results given here are generic for the potentials of the form $V \approx -\phi^{2n}$. On the other hand, greater $\mu$ indicates greater mass and radius but negative energy density at the inner layers of the configuration becomes greater as well. In other words, the configuration becomes larger in volume preserving all its features.

\subsection{Asymptotic Solutions}

Following the steps given in Refs. \cite{sotiriou,Arapoglu2019} it is possible to show that the asymptotic behavior of the system depends on the potential of the scalar field. The value of the potential (and its derivative) at spatial infinity determines the derivative of the scalar field which specifies the asymptotic character of the metric functions. If the value of the scalar field at spatial infinity $\big(\,\phi(r\!\rightarrow\!\infty)\!=\!\phi_\infty\,\big)$ has a constant limit and if this limit also satisfies $V(\phi_\infty) = 0$, then as seen from Eq.\ \autoref{eq:thesystem}, the metric functions, $f(r)$ and $g(r)$, give the Schwarzschild solution. Otherwise, the non-zero value of the potential makes the metric functions divergent and asymptotically flat solution can not be obtained. This problem occurs with Higgs-like potential for $\phi_\infty=0$ as stated before.

In order to show the above claim mathematically, it is appropriate to write the equation of motion of the scalar field, i.e. Eq.\ \autoref{eq:fieldeq2}, in the exterior region as
\begin{equation}
\begin{aligned}
	\big(1 & + 6\xi^2\kappa_{\rm eff}\,\phi^2\big) \boxempty \! \phi \\ & \qquad + \xi \big( 1+6\xi \big) \kappa_{\rm eff}\,\phi\,\nabla^c \phi \, \nabla_{\!\!c} \, \phi \\ & \hspace{3cm} = -4\xi\kappa_{\rm eff} \phi V + \der{V}{\phi} \:.
\end{aligned}
\label{eq:sca_field_pot}
\end{equation}
It seems from this form that the terms on the right-hand side determine the dynamics of the scalar field and as stated above they depend on the potential and its derivative. This implies that, in order to get a stationary solution for the scalar field outside the star, the potential of the scalar field has to satisfy a second condition at spatial infinity which is
\begin{equation}
	\bigg[\! -4\xi\kappa_{\rm eff}\,\phi\,V(\phi) + \der{\,V(\phi)}{\phi} \bigg] \bigg|_{\phi = \phi_\infty} = 0 \:.
\label{eq:second_cond}
\end{equation}
Therefore, appropriate central values for the scalar field (as the initial condition) have to be chosen such that the asymptotic value of the scalar field, $\phi_\infty$, satisfies $V(\phi_\infty) = 0$ and Eq.\ \autoref{eq:second_cond}.

The condition given by Eq.\ \autoref{eq:second_cond} for the potential that we have considered becomes 
\begin{equation}
    \phi_\infty \big(\phi_\infty - \nu\big) \big(\phi_\infty + \nu\big) = 0 \:, \quad \nu \equiv \sqrt{\sfrac{\mu^2}{\lambda + \kappa\xi\mu^2}}
\end{equation}
and it turns out that the only value which satisfies both conditions, i.e. $V(\phi_\infty)=0$ and Eq.\ \autoref{eq:second_cond}, is $\phi_\infty=0$. This asymptotic value is also the stable critical point of the effective potential as explained before.

Since the form of the potential and the initial conditions for the scalar field studied in this work obey the above conditions, the asymptotic forms of Eqs.\ \autoref{eq:components_tt} and \autoref{eq:components_rr} can be written as
\begin{equation}
	\der{g}{r} \approx \sfrac{1 - e^{2g}}{2r} \: , \qquad
	\der{f}{r} \approx \sfrac{e^{2g} - 1}{2} \: ,
\end{equation}
which have the solutions
\begin{equation}
	e^{2g(r)} \approx \bigg( 1 - \sfrac{r_o}{r} \bigg)^{\!-1} \: , \qquad
	e^{2f(r)} \approx 1 - \sfrac{r_o}{r} \: ,
\label{eq:metric_approx}
\end{equation}
where $r_o$ is the integration constant. For the scalar field, on the other hand, we plug the above expressions for the metric functions into Eq.\ \autoref{eq:scalar_eq} and consider only the leading order terms to obtain
\begin{equation}
    \phi'' + \sfrac{2}{r} \phi' + \mu^2 \phi \approx 0 \:.
\end{equation}
Then, solution to this equation is obtained as
\begin{equation}
    \phi(r) \approx \phi_{\rm c}\,\sfrac{\sin{\mu r}}{\mu r}
\label{eq:phi_approx}
\end{equation}
where $\phi_c$ is the initial value for the scalar field. However, one should note that this is also the solution for the minimal coupling case, i.e. $\xi=0$. Therefore, for the mass integral first thing to do is setting $\xi=0$ in order to keep the consistency in our approximate solutions. Then, the mass integral given in Eq.\ \autoref{eq:mass} becomes
\begin{equation}
	m(r) \approx M_\rho + 4\pi \int_{0}^{r} r^2 \bigg[ \sfrac{1}{2} \big(\phi'\big)^2 e^{-2g} -\sfrac{1}{2}\mu^2\phi^2 \bigg] \, dr
\end{equation}
with the following definition
\begin{equation}
	M_\rho = \int_{0}^{r} 4\pi\rho\,r^2\,dr \:.
\label{eq:thesolution}
\end{equation}
Plugging the solutions given in Eqs.\ \autoref{eq:metric_approx} and \autoref{eq:phi_approx} into this expression we have
\begin{equation}
\begin{aligned}
	m(r) \approx M_\rho & + 2\pi \phi_c^2 \int_{0}^{r} r^2 \Bigg\{ \bigg( \sfrac{\cos \mu r}{r} - \sfrac{\sin \mu r}{\mu r^2} \bigg)^{\!2} \\ & \qquad\qquad \times \bigg( 1 - \sfrac{r_o}{r} \bigg) - \sfrac{\sin^2\! \mu r}{r^2} \Bigg\} \, dr
\end{aligned}
\end{equation}
and, furthermore, neglecting the convergent terms we get the final result as
\begin{equation}
	m(r) \approx M_\rho + \sfrac{\pi \phi_c^2}{\mu} \sin 2\mu r
\end{equation}
which explains the oscillatory behavior of the mass function outside the configuration as shown in Fig.\ \autoref{fig:quarky_funcs}.

\section{CONCLUSION} \label{conclusion}
In this work we have investigated the structure of spherically symmetric and static configurations modelled by realistic EOS for neutron matter in the presence of a scalar field having potential \autoref{eq:potential}. By analyzing the whole system numerically we have given the resulting M-R relations for a sample of central value of the scalar field with two different values of the coupling constant. We have also examined the radial profiles of the scalar field, densities and the mass function for a specific configuration.

As mentioned in the previous sections the scalar field with potential \autoref{eq:potential} describes three different types of solution depending on the initial conditions and the value of the potential parameters. We have chosen a solution type corresponding to a relatively wide range of initial conditions for the scalar field which provides asymptotic flatness in the context of the metric functions but causes periodic oscillations in the mass function due to the fact that the radial profile of the scalar field does not obey an exponential suppression outside the star. However, putting aside this problem due to its ineffectiveness on the determination method of the radius and the mass of the star as explained in the previous sections, we have examined the consequences of that particular solution on the inner structure of the configuration. On the other hand, in the central region the scalar field causes the total radial density of the star to take negative values which are necessary to obtain a stable wormhole solution \cite{Thorne}. Moreover, although different types of scalar fields have been considered, there are studies that model the presence of a wormhole inside a stellar configuration \cite{Dzhunushaliev2,Dzhunushaliev1}. But in order to reconcile those models with the results of this paper requires further investigation.

We have shown that the solution type considered in this paper causes neutron star M-R curves to alter drastically. Although we have imposed EOS for neutron matter to the system, resulting M-R relations are similar to the objects known as SQS. This is independent of the coupling of the scalar field to the gravitational sector. We have also shown that it is possible to get this relations for both bare SQS and SQS with nuclear crust depending on the value of the parameters of the model. Moreover, in case of the minimal coupling, there is only one free parameter, namely the central value for the scalar field, which can also be restricted through the mass and the radius observations as illustrated in Fig.\ \autoref{fig:mr_curves}.

In the original symmetron paper \cite{Hinterbichler2010} it was shown that in order to satisfy the local tests of gravity it is enough to consider that the Milky Way is screened. However, a general description without any assumption brings about two other possibilities, namely unscreened and partially screened objects as stated in Ref. \cite{Sakstein2015}. These cases may arise due to chosen values of the parameters in the model and/or inadequate radial size of the objects which do not allow the scalar field to take its vacuum expectation value outside the star. Although we formulate the system in the Jordan frame, it is clear that how the scalar field can cause a radical change in M-R relations of neutron stars. It is noteworthy to point out that we have obtained the same results in the minimally coupled case that does not show any deviations in the constraints determined by the parameterized post-Newtonian formalism. For future works, on the other hand, the cases that correspond to partially screened objects could be investigated in order to check whether it is possible to eliminate the complications occurred here due to behavior of the scalar field in the central and the exterior region of the star.

\bibliographystyle{apsrev4-1}
\bibliography{references}

\end{document}